\documentclass{epl}

\usepackage{dcolumn}% Align table columns on decimal point
\usepackage{bm}% bold math
\usepackage{amsmath}
\usepackage{graphicx}

\title{Unanimity Rule on networks}

\author{Renaud Lambiotte\inst{1}, Stefan Thurner\inst{2} and Rudolf Hanel\inst{3,2} }

\institute{
\inst{1} Universit\'e de Li\`ege, Sart-Tilman, B-4000 Li\`ege, Belgium\\
\inst{2} Complex Systems Research Group; HNO; Medical University of Vienna; W\"ahringer G\"urtel 18-20; A-1090; Austria\\
\inst{3} Institute of Physics; University of Antwerp; Groenenborgerlaan171; 
2020 Antwerp; Belgium 
}

\begin{document}

\maketitle

\begin{abstract}
We introduce a model for innovation-, evolution-  and opinion dynamics whose spreading 
is dictated by unanimity rules, i.e. a node will change its (binary) state only if all of its neighbours 
have the same corresponding  state. It is shown that a transition takes place depending on 
the initial condition of the problem. In particular, a critical number of initially activated nodes 
is needed so that the whole system gets activated in the long-time limit. The influence of the 
degree distribution of the nodes is naturally taken into account.  For simple 
network topologies we solve the model analytically, the cases of random, small-world and 
scale-free are studied in detail.  

%\noindent
%Keywords: Evolution, innovation, opinon dynamics, phase transition, food chains

\end{abstract}

\section{Introduction}

In general, the discovery or emergence  of something  depends on the combination
of several parameters, all of them having to be  {\em simultaneously} met.  One may think of 
economy, where the production of a good depends on the production or existence  of other goods 
(e.g. to produce a car one needs the wheel, the motor and some fioritures). In return, 
this new discovery  opens new possibilities and needs that will lead to the production 
of yet new goods (e.g. the simultaneous existence of the car and of alcohol directly leads 
to the invention of the air bag). This auto-catalytic process is a very general process  
\cite{hanel,farmer,kauffman,stadler} and obviously applies to many situations not only related to 
innovation, but also to evolution, opinion formation, food chains etc. 
One may even think of  the dynamics of scientific ideas, music genres, or any other field where 
the emergence of a new element possibly leads to new combinations and new elements. 
This feedback is responsible for the potential explosion of the number of items, such as observed 
e.g. in the Cambrian explosion).  This ''explosion'' has been shown to be identical of a phase transition 
in a Van der Waals gas \cite{hanel}. 
After mapping the above catalytic reactions onto a network structure, where nodes represent 
items and directed links show which items are necessary for the production of others, it is tempting 
to introduce a unanimity rule (UR): a node on the network is activated only if all the nodes arriving 
to it through a link are activated. Surprisingly, the dynamics of such an unanimity rule, 
that is a straightforward generalization of the usual majority rules of opinion dynamics 
\cite{galam,sznajd,redner,voter2,voter3}, is poorly known \cite{hanel}. 
In this Letter, we are interested on the spreading dynamics of the UR and try to understand which 
conditions have to be satisfied so that the network gets fully activated in the long time limit. 
 
 \begin{figure}
\includegraphics[width=1.8in]{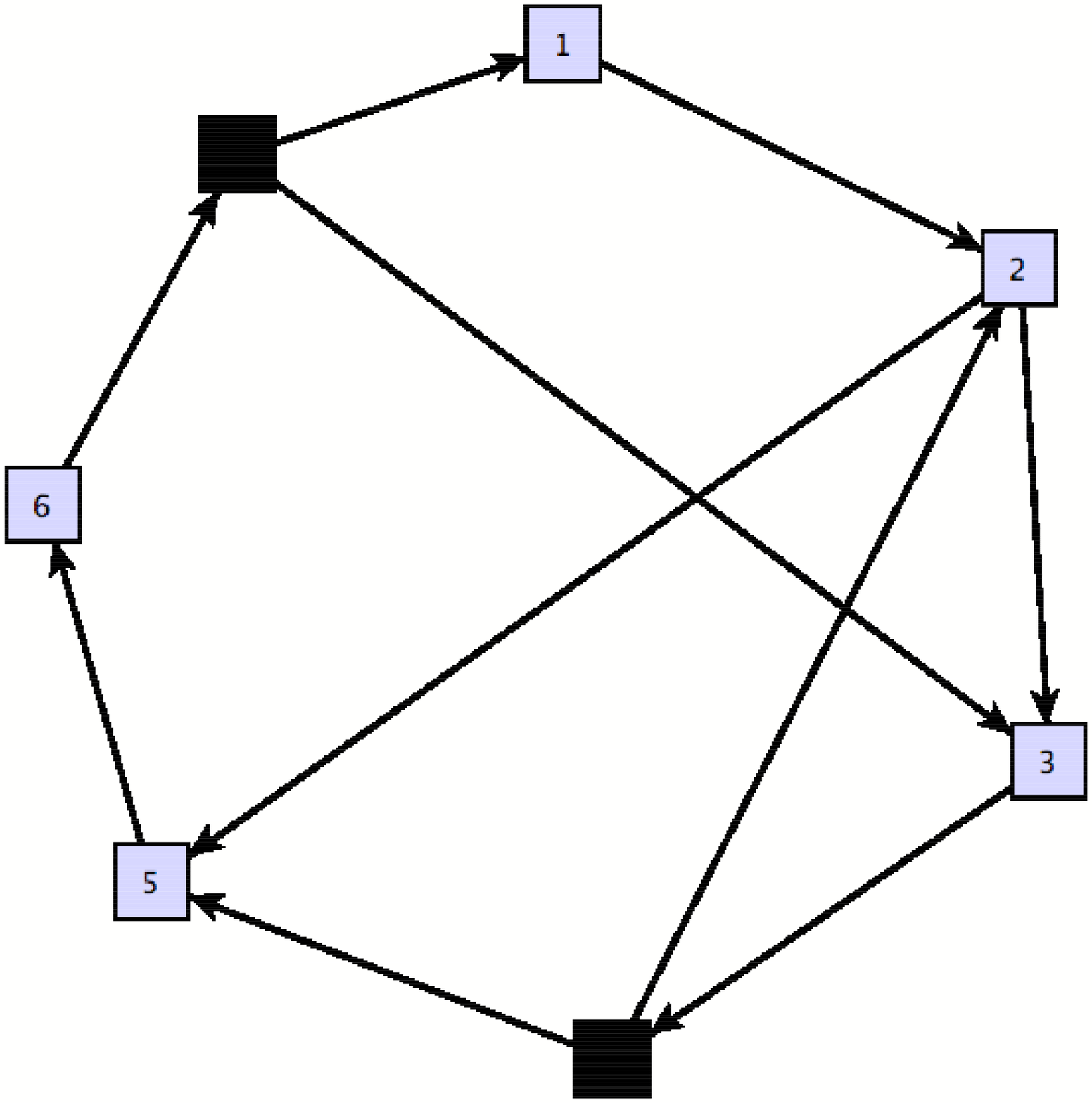}
\includegraphics[width=1.8in]{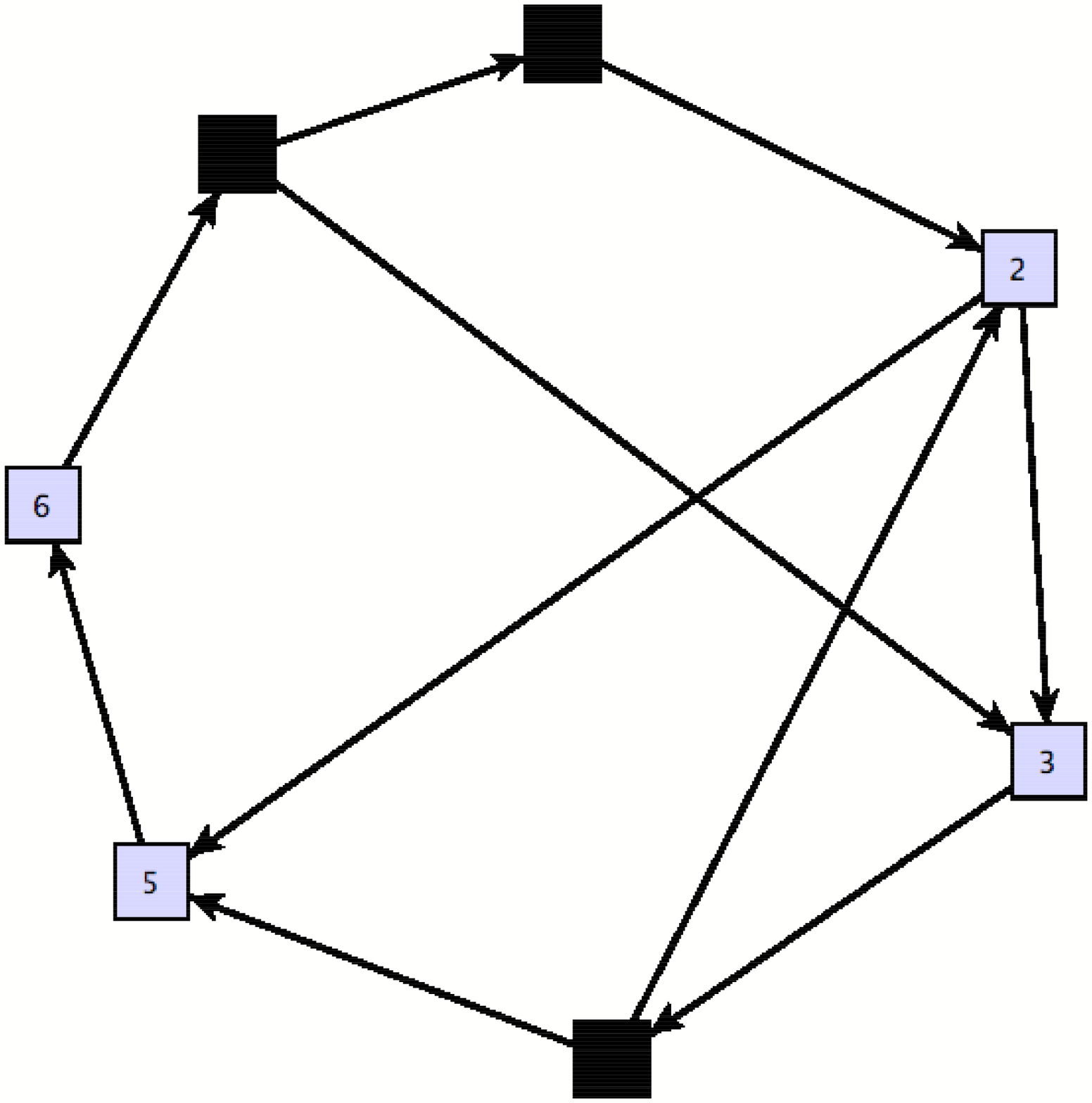}
\includegraphics[width=1.8in]{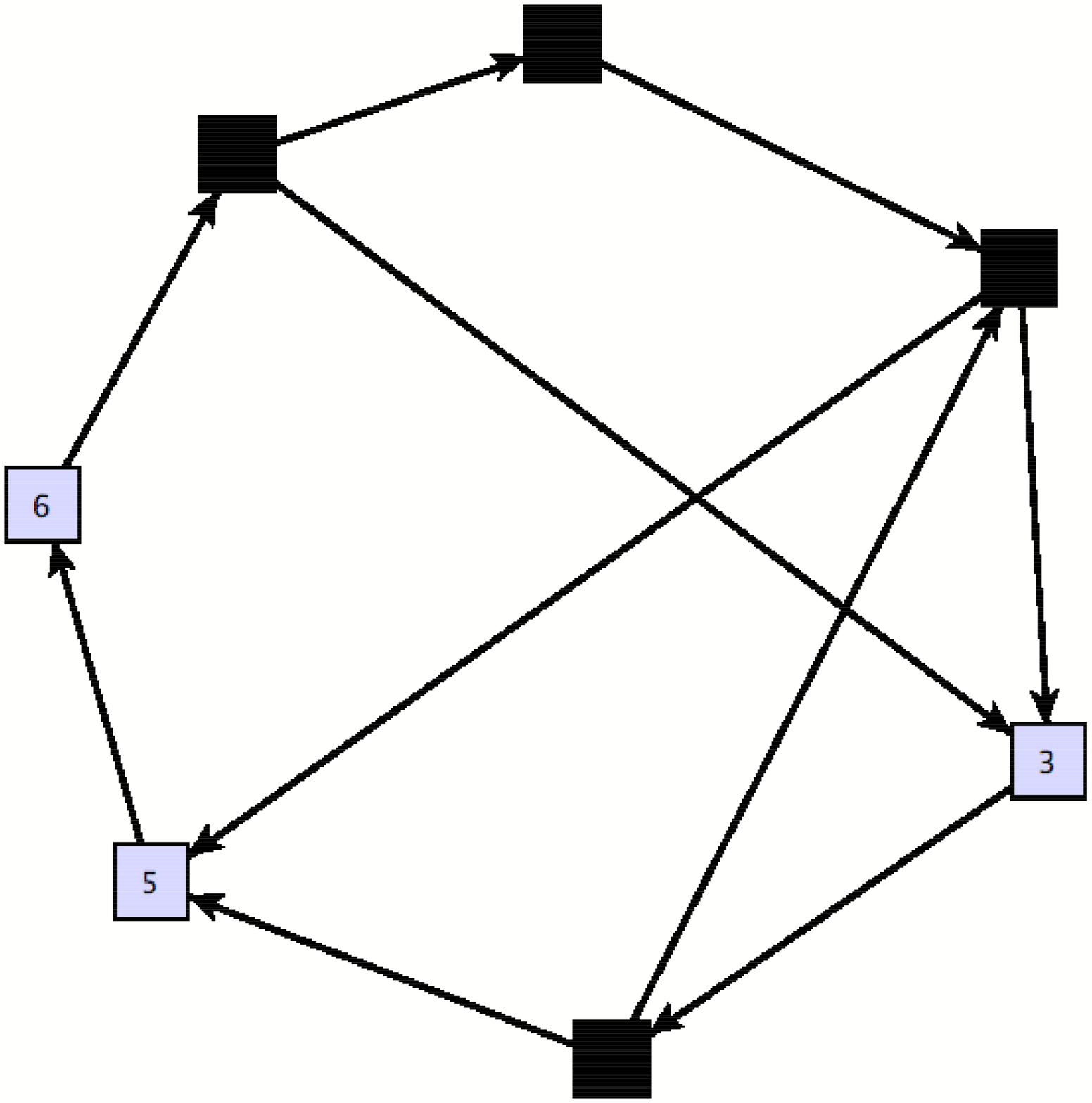}

\caption{First two steps of UR starting from an initial network of 7 nodes, 2 of them being activated. Initially 
there is only one node among the non-activated nodes that satisfies the unanimity rule. It gets therefore 
activated at the first time step. At that time, there is a new node whose 2 incoming links come from activated nodes. 
It gets  activated at the second time step. It is straightforward to show that this system gets fully activated at 
the fourth time step.}
\label{fig1}      
\end{figure}

\section{Unanimity model} 
%Let us now introduce the model in detail.  
The network is composed of $N$ nodes related through directed links. Each node exists in one of two states: 
activated or inactivated.  The  number of nodes with indegree $i$ (the indegree of a node is defined to be the 
number of links arriving to it) is denoted by $N_i$ and depends on the underlying network structure. 
Initially (at $t=0$) there are  $A_0$ nodes which are activated, among which  $A_{i,0}$ have an 
indegree $i$.
% ($\sum_i A_{i,0}=A_0$). 
In general, the number of nodes of type $i$, activated at time $t$, is $A_{i,t}$. It is also useful to introduce 
the quantities  $n_i=\frac{N_i}{N}$ and $a_{i,t} = \frac{A_{i,t}}{N_i}$ which are  the proportions of nodes of 
type $i$ in the network (indegree distribution) and the probability that such a node $i$ is activated, respectively. 
$a_{t} = \frac{A_{t}}{N}=\sum_{i} n_i a_{i,t}$ is the fraction of activated nodes in the whole network at time $t$.
The unanimity rule is defined as follows (see Fig.1). At each time step, each node is considered. 
If all the links arriving to a specific unactivated node $i$ originate at nodes which are activated at $t-1$, 
$i$ gets activated at $t$. Otherwise, it remains unactivated. 
The process is applied iteratively until the system reaches a stationary state, characterized by an 
asymptotic value $a_\infty$. 
In the following, we are interested in the relation between $a_\infty$ and $a_0$, 
i.e.  what is the final occupation of the network as a function of its initial occupation on a specific network. 
Let us mention the fact that each node may be produced by only {\em one} combination of 
(potentially many, depending on the indegree) nodes.  
This is a modification of the model of Hanel et al. \cite{hanel}, where more than one pairs of (two) nodes could 
produce new elements and  will lead to a different equation for the activation evolution, as shown below. 
The dynamics studied here implies that nodes with a higher indegree will be activated with a probability 
smaller than those with a smaller indegree (because the former have more conditions to be fulfilled). 

\section{Master equation} Let us now derive an evolution equation for $A_{i,t}$ and $A_t$. 
To do so it is helpful to consider the first time step and than to iterate. There are initially 
$A_0$ activated nodes, $A_{i,0}=A_0 N_i / N$ of them being of indegree $i$ on average 
(the activated nodes are randomly chosen in the beginning). The ensemble of  $A_{i,0}$ 
nodes is called the initial set of indegree $i$. By construction, the probability that $i$ 
randomly chosen nodes are activated, is $a_0^i$ ($i$ is an exponent). Consequently, the average number of 
nodes with indegree $i$ and who respect the unanimity rule is $N_i a_0^i$ while the number 
of such nodes that are not yet occupied is
\begin{eqnarray}
\label{delta0}
\Delta_{i,0} = (N_i - N_i a_0) a_0^i   , 
\end{eqnarray}
and, on average, the total number of occupied nodes with indegree $i$ evolves as:
\begin{eqnarray}
A_{i,1} = A_{i,0} + \Delta_{i,0}.
\end{eqnarray}
Let us stress that we have implicitely assumed that there are no indegree correlations between 
neighboring nodes in order to derive Eq.\ref{delta0}. At the next time step, the average number 
of nodes with indegree $i$, who respect the unanimity rule and who are outside the initial set is 
$(N_i- N_i a_0) a_1^i$. Among those nodes, $\Delta_{0,i}$ have already been activated during the first 
time step, so that the average number of nodes who get activated at the second time step is:
 \begin{eqnarray}
\label{delta1}
\Delta_{i,1} = (N_i - N_i a_0) (a_1^i - a_0^i) .
\end{eqnarray}
Note that Eq.\ref{delta1} is valid because no node in $\Delta_{i,1}$ also belongs to $\Delta_{i,0}$. 
This is due to the fact that each node can only be activated by {\em one} combination of $i$ 
nodes in our model, so that no redundancy is possible between $\Delta_{i,1}$ and $\Delta_{i,0}$. 
By proceeding similarly, it is straightforward to show that the contributions $\Delta_{i,t}$ read
\begin{eqnarray}
 \label{deltai} 
 \Delta_{i,t} = (N_i - N_i a_0) (a_t^i - a_{t-1}^i) ,
\end{eqnarray}
with $a_{-1}=0$ by convention. The number of activated nodes evolve as
\begin{eqnarray}
A_{i,t+1} = A_{i,t} + \Delta_{i,t}.
\end{eqnarray}
By dividing by $N_i$, one gets a set of equations for the proportion of nodes $a_i \in [0,1]$:
\begin{eqnarray}
 a_{i,t+1} = a_{i,t} + (1-a_{0}) (a_{t}^i - a_{t-1}^i ),
\end{eqnarray}
where the coupling between the different proportions $a_{i,t}$ occurs through the 
average value $a_t=\sum_i n_i a_{i,t}$, as defined above. Finally, by multiplying by 
the indegree distribution $n_i$ and summing 
over all values of $i$, one gets a closed equation for the average proportion of activated nodes 
in the network that reads
\begin{eqnarray}
\label{equation}
  a_{t+1} = a_{t} + (1-a_{0}) \sum _i n_i (a_{t}^i - a_{t-1}^i ).
\end{eqnarray}
Let us stress that Eq.\ref{equation} is non-linear as soon as $N_i\neq0$, $i>1$. Moreover, 
it is characterized by the non-trivial presence of the initial condition $a_0$ in the right hand 
non-linear term and is therefore highly non-local in time. Eq.\ref{equation} 
explicitly shows how the indegree distribution $n_i$, affects the propagation of activated 
nodes in the system.

\section{Theoretical results} 
In this section, we focus on simple choices of $n_i$ in order to apprehend analytically 
the behavior of Eq.\ref{equation}. The simplest case is $n_1=1$ for which Eq.\ref{equation} reads
\begin{eqnarray}
a_{t+1} = a_{t} + (1-a_{0})  (a_{t} - a_{t-1} ).
\end{eqnarray}
This equation is  solved by recurrence:
\begin{eqnarray}
a_{1} &=& a_{0} + (1-a_{0})  a_0 \cr
a_{2} &=& a_{0} + (1-a_{0})  a_0 + (1-a_0) [a_{0}+(1-a_{0})  a_0 - a_0]\cr
&=&a_{0} + (1-a_{0})  a_0 + (1-a_{0})^2  a_0
\end{eqnarray}
and in general
\begin{eqnarray}
a_{t} =  \sum_{u=0}^{t}  (1-a_0)^u  a_0=   1-(1-a_0)^{t+1}. 
\end{eqnarray}
This last expression is easily verified: 
\begin{eqnarray}
a_{t+1} &=&  \sum_{u=0}^t  (1-a_0)^u a_0 + (1-a_{0})  
 ( \sum_{u=0}^t   (1-a_0)^u a_0 -  \sum_{u=0}^{t-1}   (1-a_0)^u a_0)\cr
 &=&   \sum_{u=0}^t  (1-a_0)^u a_0 +    (1-a_0)^{t+1} a_0 = \sum_{u=0}^{t+1}  (1-a_0)^u  a_0 .
\end{eqnarray}
The above solution implies that any initial condition converges toward the asymptotic state $a_\infty=1$, i.e. whatever the initial condition, the system is fully activated in the long time limit. The relaxation to $a_\infty=1$ is exponentially fast $\sim e^{\ln (1-a_0) t}$.

\begin{figure}
\includegraphics[width=2.7in]{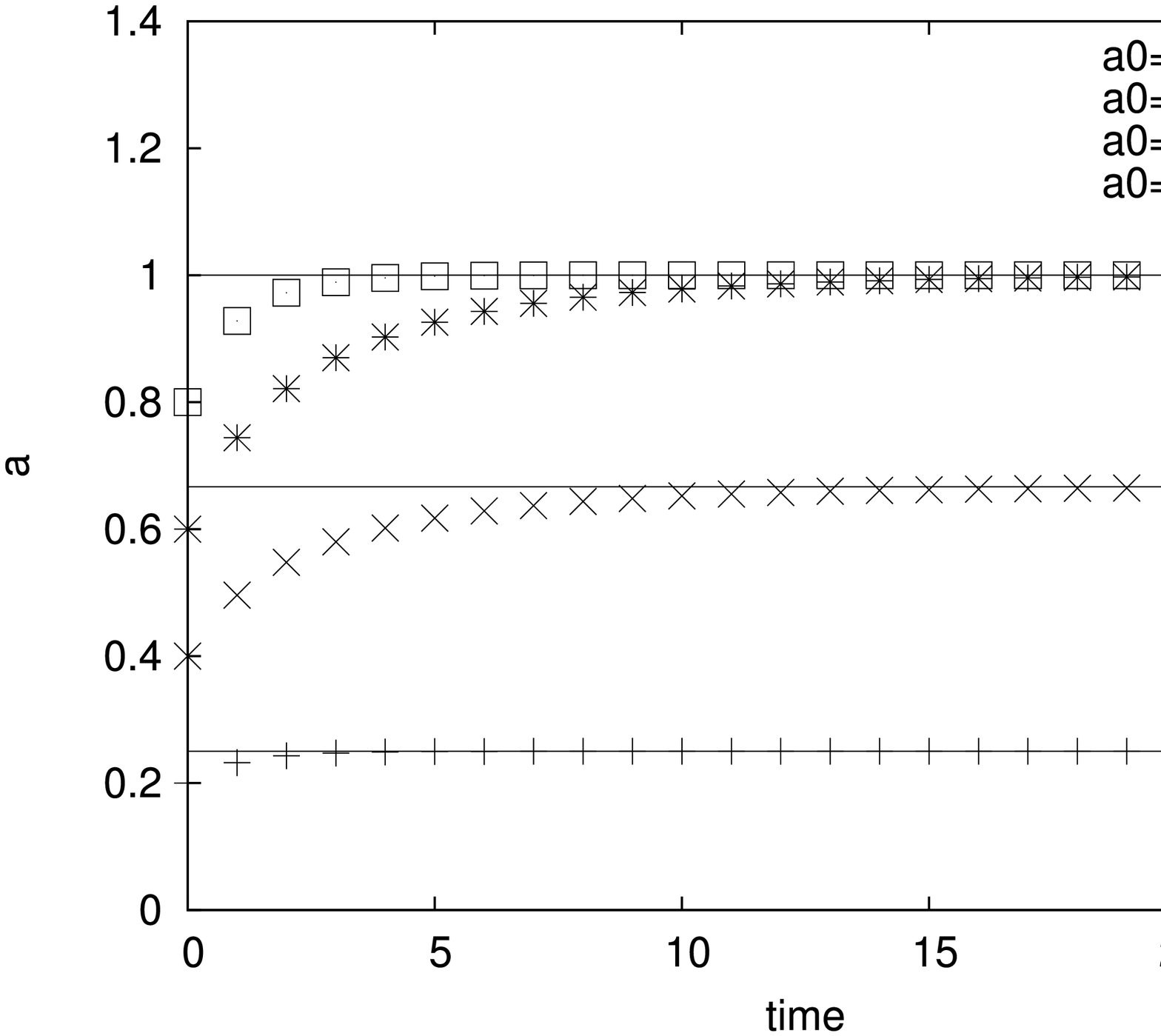}
\includegraphics[width=2.7in]{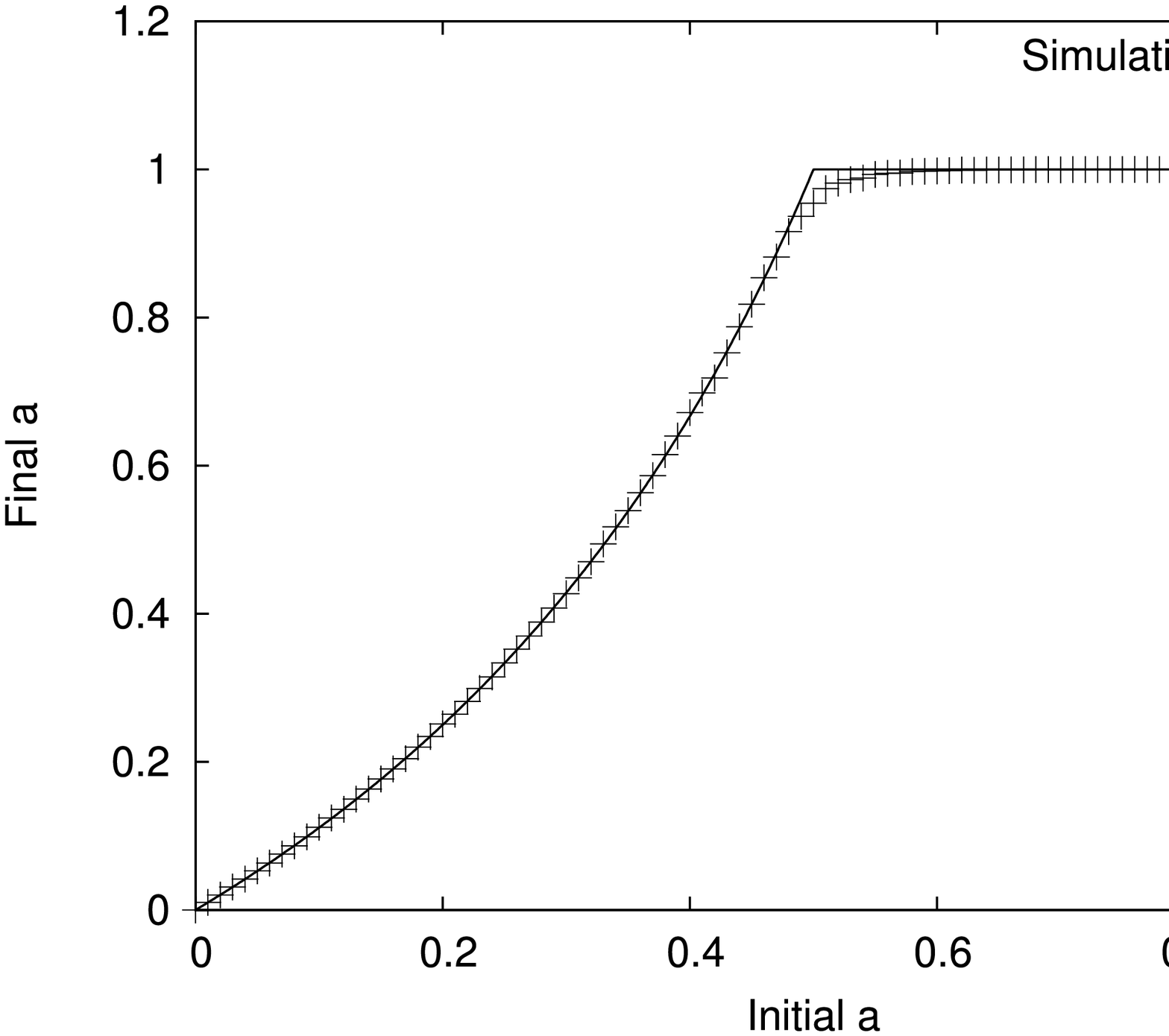}

\caption{Numerical integration of Eq.\ref{order2}. In the left figure, we plot the time evolution of $a_t$, together with the theoretical prediction \ref{solution}. In the right figure, we plot the relation $a_\infty(a_0)$ that obviously shows a transition at $a_0=1/2$. Eq.\ref{solution} is in perfect agreement with the numerical integration of Eq.\ref{order2} (line). 
The agreement with the numerical simulations of the model itself is also excellent.}
\label{fig2}      
\end{figure}

Let us now focus on the more challenging case $n_2=1$ where all the nodes have an 
indegree of 2 by construction. In that case, Eq. \ref{equation} reads
\begin{eqnarray}
\label{order2}
a_{t+1} = a_{t} + (1-a_{0})  (a_{t}^2 - a_{t-1}^2 ).
\end{eqnarray}
The non-linear term does not allow to find a simple recurrence expression as above. 
Though, a numerical integration of Eq.\ref{order2} (by using Mathematica for instance) 
shows that  the leading terms in the Taylor expansion of $a_{t}$ behave like
\begin{eqnarray}
  a_{t} = \sum_{i=1}^{t+1}  a_0^i  +  {\cal O} (t+2)
\end{eqnarray}
thus suggesting that the asymptotic solution is 
\begin{eqnarray}
  \label{solution}
  a_\infty = \frac{a_0}{1-a_0}.
\end{eqnarray}
This solution should satisfy the normalization constraint $a_\infty \leq 1$, so that it can hold 
only for initial conditions $a_0>1/2$. This argument suggest that a transition takes place 
at $a_c=1/2$, such that only a fraction of the whole system gets activated when $a_0<a_c$ 
while the whole system activates above this value (see Fig.2). 
We verify the approximate solution Eq.\ref{solution} by  
looking  for a solution of the form $a_t=\frac{a_0}{1-a_0} (1+ \epsilon_t)$. 
By insterting this expression into Eq.\ref{order2}, one gets the recurrence relations:
\begin{eqnarray}
\epsilon_{t+1} &=&  \epsilon_t +  a_0 (1+\epsilon_t )^2 - a_0 (1+\epsilon_{t-1} )^2 \cr
\epsilon_{t+1} &=&  \epsilon_t +  2 a_0 (\epsilon_t - \epsilon_{t-1} ), 
\end{eqnarray}
where the second line is obtained by keeping only first order corrections in $\epsilon$. In the continuous time limit, 
keeping terms until the second time derivative, one obtains
\begin{eqnarray}
(1-2 a_0)  \partial_t  \epsilon_t + 1/2 (1+2 a_0) \partial_t^2  \epsilon_t =0 ,
\end{eqnarray}
whose exponential solutions read  $\epsilon_t =e^{- \lambda t}$ with
\begin{eqnarray}
\lambda=  \frac{1}{2}\frac{(1-2 a_0)}{(1+2 a_0)}.
\end{eqnarray}
This is a relaxation to the stationary state $a_\infty$ only when $a_0<1/2$, thereby confirming a 
qualitative change at $a_c=1/2$.

\begin{figure}
\includegraphics[width=2.7in]{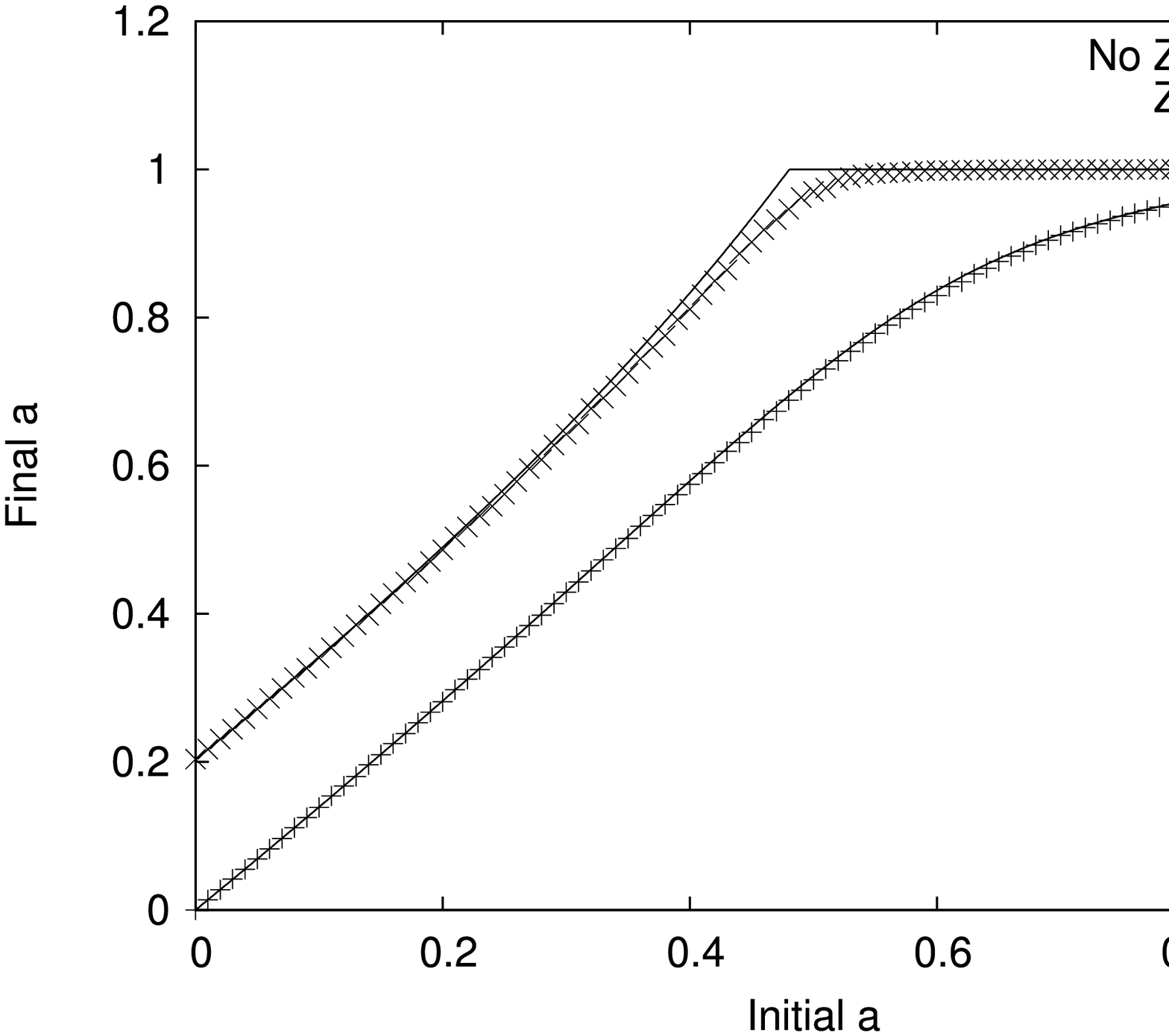}
\includegraphics[width=2.7in]{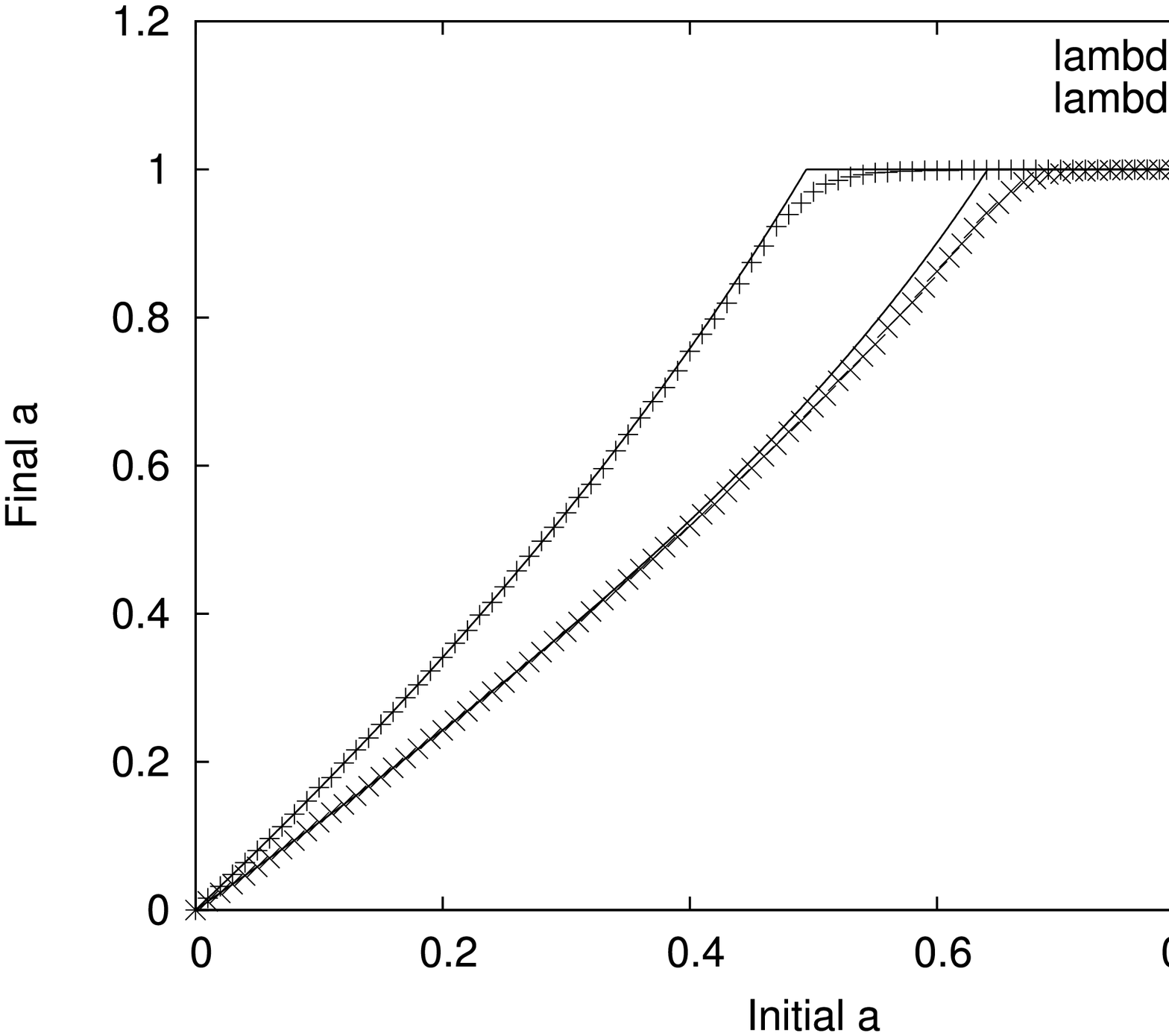}

\caption{$a_\infty(a_0)$ obtained for random networks with $L=2 N$ (left figure). 
The measured indegree distribution is $n_0=0.14$, $n_1=0.27$, $n_2=0.27$... 
The No Zero and Zero versions (see text)are shown. In the right figure, $a_\infty(a_0)$  
for small-world networks $L=\lambda N$ short-cuts. When $\lambda=1$, the measured indegree 
distribution is $n_0=0.0$, $n_1=0.37$, $n_2=0.37$, etc. When $\lambda=2$, it is $n_0=0.0$, 
$n_1=0.14$, $n_2=0.27$... In both figures, the total number of nodes is $N=10000$ 
and the results are averaged over $100$ realizations of the process. The solid lines are 
the corresponding numerical solutions of Eq.\ref{equationZero}, evaluated with the 
empirical values $n_i$.}
\label{fig3}      
\end{figure}

\section{Some network topologies}
Let us now focus on more reasonable topologies and compare the results obtained from 
Eq. \ref{equation} with numerical simulations of the UR. We focus on three 
types of networks, purely random networks \cite{Erdos}, small-world like networks \cite{watts} 
and Barabasi-Albert networks  \cite{bar} (growing networks with preferential attachment). 
The excellent agreement with Eq. \ref{equation} suggest that the formalism should apply to 
more general situations as well. The random network  was obtained by randomly assigning $L$ directed 
links over $N$ nodes. The small-world network was obtained by starting from a directed ring 
configuration  and than randomly assigning $L$ directed links (short-cuts) over the nodes, 
i.e. the total number of links in that case is $L+N$ (The network drawn in Fig.1 is such network 
with $N=7$ nodes and $L=3$ short-cuts). Let us note that the small-world network can be 
viewed as a food chain with a well-defined hierarchy between species together with some random 
short-cuts. In that case, UR can be interpreted as an extinction model (if all the species that 
one species eats go extinct, this species will also go extinct). The Barabasi-Albert network was built 
starting from one seed node and adding nodes one at a time until the system is composed of 
$N$ nodes. At each step, the node first connects to a  randomly chosen node and, with 
probability $p$, it re-directs its link to the father of selected node. This method is well-known 
to be equivalent to preferential attachment and to lead to the formation of fat tail degree 
distributions $k^{\nu}$, with $\nu=1+1/p$, \cite{krapi}.

\begin{figure}
\includegraphics[width=2.7in]{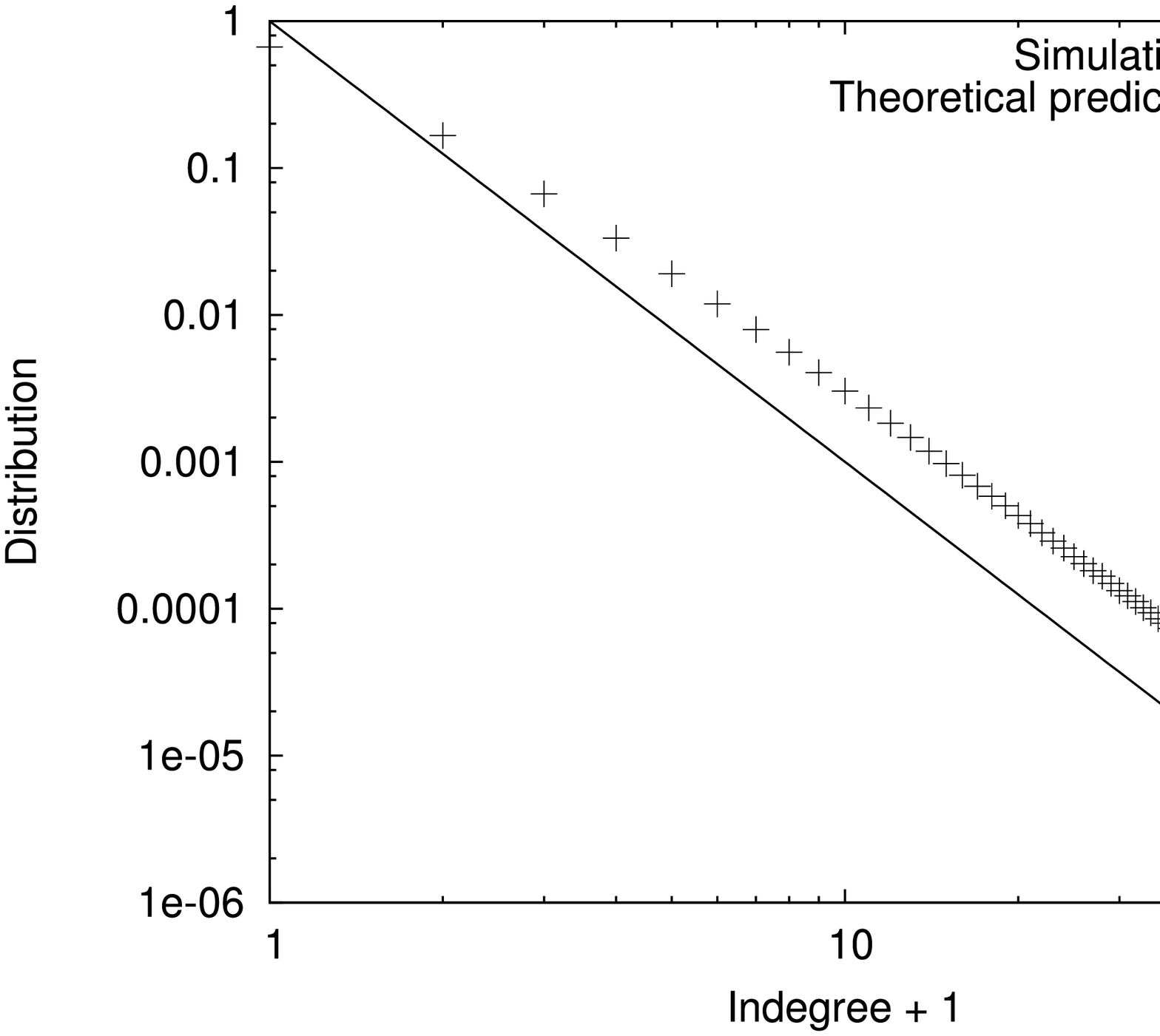}
\includegraphics[width=2.7in]{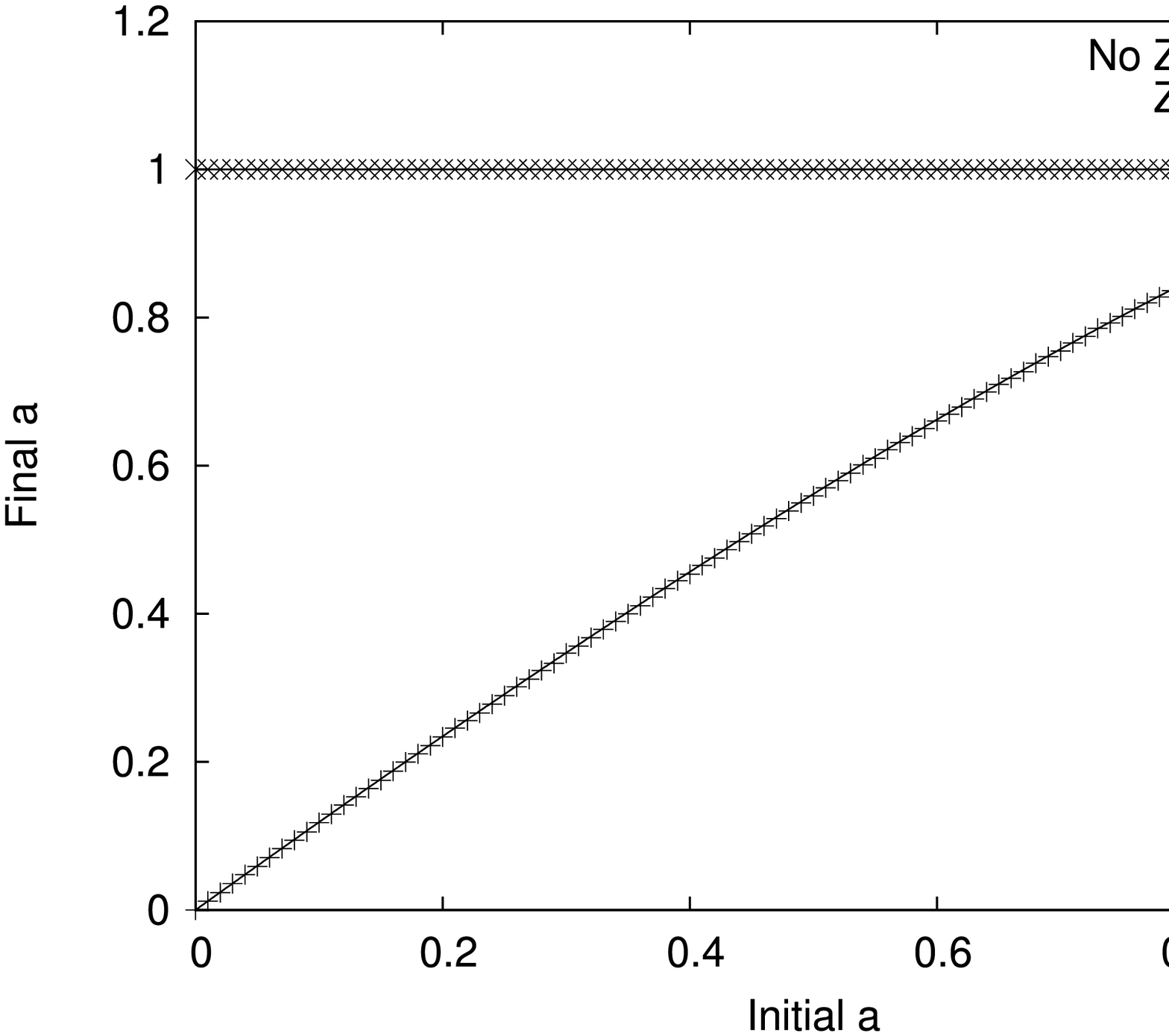}

\caption{(left figure) distribution of indegrees for the scale-free network, 
$m_i=n_{i+1}$ (this shift being made to plot $n_0$ in log-log scale). 
The solid line is the theoretical asymptotical prediction $k^{-(1+1/p)}$, for $p=1/2$. 
The total number of nodes 
is $N=10000$, results are averaged over $100$ realizations of the process. 
(right figure) $a_\infty(a_0)$  for the same network. In the No Zero version, one observes that the all the nodes 
are finally activated whatever the initial condition. The solid lines correspond to 
numerical values of Eq.\ref{equationZero} for the empirical values $n_i$.
}
\label{fig4}      
\end{figure}

Once the underlying network is built, we randomly assign $a_0 N$ active nodes to the network and apply the 
unanimity rule. The evolution  stops once a stationary state is reached. The asymptotic value $a_\infty$ is 
averaged over several realizations of the process (on several realizations of the underlying network). 
In the small-world network, each node receives at least one incoming link. This is not the case for the random- 
or the BA networks, for which one has to discuss the   ambiguous dynamics of nodes with zero incoming links. 
Two choices are possible. Either these nodes can not be activated in the course of time, because they are not 
reached by any other node (No Zero version), or all of them are get activated at the first time step, 
thereby assuming that their activation does not require any {\em first knowledge}  (Zero version). 
The choise is a question of interpretation. The two versions are associated to different evolution equations:

\begin{eqnarray}
\label{equationZero}
a_{t+1} &=& a_{t} + (1-a_{0}) \sum _{i=1}^{\infty} n_i (a_{t}^i - a_{t-1}^i )  ~~{\rm No} ~{\rm Zero}~{\rm version}  \cr
a_{t+1} &=& a_{t} + (1-a_{0}) \sum _{i=0}^{\infty} n_i (a_{t}^i - a_{t-1}^i ) ~~{\rm Zero}~{\rm version}  ,  
\end{eqnarray}
and leads to quite different behaviors (Figs.3 and 4).
In the case of small-world networks, the above equations are obviously equivalent.
To compare the simulation results with Eq.\ref{equationZero}, we 
also measure the indegree distributions  of the networks $n_i$ generated during 
the simulations and integrate Eq.\ref{equationZero} with these empirical values. 
The agreement is excellent, except close to the transition points where finite size 
effects are expected. It is worth noting that the importance of nodes with zero incoming 
links is much higher in (growing) Barabasi-Albert like networks (Fig.4), so that the 
difference between the two versions is quite pronounced, as expected. 
Let us also mention that Eq.14 has been successfully verified for a random network 
where the indegree of each node is exactly 2 ($n_2=1$), as shown in Fig.2.

\section{Discussion}
In this Letter, we have introduced a simple model for innovation whose dynamics is based on Unanimity Rule. 
It is shown that the discovery of new items on the underlying network opens perspectives for the 
discovery of new items. This feedback effect may lead to complex spreading properties, embodied by the existence of a {\em critical size} for the initial activation, that is necessary for the complete activation of the network in the long time limit. The problem has been analyzed empirically on a large variety of network structures and has been successfully described by recurrence relations for the average activation. Let us stress that these recurrence relations have a quite atypical form, due to their explicit dependence on the initial conditions. Moreover, their non-linearity makes them a hard problem to solve in general. Finally, let us insist on the fact that Unanmity Rule is a general mechanism that should apply to numerous situations related to innovation, opinion dynamics or even species/population dynamics.  To be consistent with our own work, we also hope that this paper will trigger the reader's curiosity and,  possibly, open new perspectives or research directions...

 {\bf Acknowledgements}
This collaboration was made possible by a COST-P10 short term mission. 
R.L. has been supported by European Commission Project 
CREEN FP6-2003-NEST-Path-012864. 
S.T. is grateful to Austrian Science Foundation projects P17621 and P19132.

\end{document}